\documentclass[acmsmall]{acmart}
\usepackage{multirow}
\usepackage{xcolor}
\usepackage{framed}
\usepackage{fancyvrb}
\usepackage{soul}
\usepackage{siunitx}
\soulregister\ref7
\soulregister\eqref7
\soulregister\cite7
\soulregister\cref7
\soulregister\Cref7
\soulregister\autoref7

\usepackage{appendix}
\usepackage{amsmath}

\colorlet{shadecolor}{blue!10!white}

\AtBeginDocument{%
  }

\setcopyright{cc}
\setcctype{by}
\acmDOI{10.1145/3808218}
\acmYear{2026}
\acmJournal{PACMSE}
\acmVolume{3}
\acmNumber{FSE}
\acmArticle{FSE211}
\acmMonth{7}
\acmSubmissionID{fse26mainb-p3803-p}
\received{2025-09-12}
\received[accepted]{2026-03-24}

\begin{document}

\title{Event-B Agent: Towards LLM Agent for Formal Model Synthesis and Repair}

\author{Hongshu Wang}
\orcid{0009-0006-0198-148X}
\affiliation{%
  \institution{National University of Singapore}
  \city{Singapore}
  \country{Singapore}
}
\email{hongshu.wang@u.nus.edu}
\authornote{Co-corresponding authors.}

\author{Xinyue Zuo}
\orcid{0009-0008-4411-3054}
\affiliation{%
  \institution{National University of Singapore}
  \city{Singapore}
  \country{Singapore}
}
\email{zuoxy@nus.edu.sg}

\author{Yuhan Sun}
\orcid{0009-0005-9132-5349}
\affiliation{%
  \institution{East China Normal University}
  \city{Shanghai}
  \country{China}
}
\email{51285902023@stu.ecnu.edu.cn}

\author{Qin Li}
\orcid{0000-0001-7476-4079}
\affiliation{%
  \institution{East China Normal University}
  \city{Shanghai}
  \country{China}
}
\email{qli@sei.ecnu.edu.cn}
\authornotemark[1]

\author{Yamine Ait Ameur}
\orcid{0000-0003-4582-9712}
\affiliation{%
  \institution{IRIT - National Polytechnic Institute of Toulouse}
  \city{Toulouse}
  \country{France}
}
\email{yamine.aitameur@toulouse-inp.fr}

\author{Jin Song Dong}
\orcid{0000-0002-6512-8326}
\affiliation{%
  \institution{National University of Singapore}
  \city{Singapore}
  \country{Singapore}
}
\email{dcsdjs@nus.edu.sg}


\begin{abstract}
  Building software that is correct by construction is a long-standing goal in software engineering, as it ensures reliability during design and development rather than after deployment. Formal methods realize this vision by enabling the expression of system behavior and requirements in mathematics, thereby guaranteeing correctness through formal verification, including theorem proving and model checking. However, the steep learning curve and demand for mathematical expertise hinder the widespread adoption of formal methods. Large language models (LLMs) have recently shown promise in bridging this gap through autoformalization. However, existing LLM-based approaches are largely limited to isolated tasks, such as theorem proving without formalization or model synthesis with insufficient verification. While valuable, these efforts do not fully exploit the potential of a more comprehensive framework in which models and proofs evolve together, a process that closely reflects real-world development practice. To address this gap, we propose Event-B Agent, a novel framework inspired by the interleaved nature of software design. Given natural language requirements, Event-B Agent constructs an initial model and iteratively repairs and refines it using formal verification feedback. Refinement simplifies proof discharge, while repair of models and proofs ensures the soundness of each refinement step. Together, these two components reinforce each other to progressively improve the model quality. Evaluation across systems of varying complexity demonstrates that Event-B Agent substantially outperforms baselines in end-to-end formal model synthesis and repair, while maintaining reasonable efficiency. These results suggest that Event-B Agent is a promising step toward correct-by-construction formal model synthesis and repair.
\end{abstract}



\maketitle

\section{Introduction} \label{sec:intro}

\emph{Correct-by-construction} model development is an important paradigm in formal methods, aiming to ensure correctness during development through formal verification rather than post-hoc testing.
Typically, requirements are incrementally formalized into a sequence of models, where each intermediate model must be verified before further refinement.
However, the steep learning curve and reliance on mathematical expertise have hindered their widespread adoption.
As system complexity increases, the number of proof obligations grows rapidly.
Representative frameworks such as the B-Method \cite{abrial1996b} and Event-B \cite{abrial2010modeling} often generate hundreds of proof obligations even for moderately complex models.
Traditionally, the iterative process of constructing formal models, discharging proof obligations, and repairing accordingly is performed manually, making it time-consuming and dependent on human expertise.

Meanwhile, the rise of large language models (LLMs) \cite{achiam2023gpt, openai_gpt5} has attracted growing attention across research communities \cite{jiang2024survey, hou2024large, zhang2025position}. 
Their natural language understanding capabilities open new opportunities to automate formalization. 
Recent research in formal methods has explored the potential of LLMs in a variety of autoformalization tasks, including linear temporal logic synthesis~\cite{cosler2023nl2spec, fuggitti2023nl2ltl} and program specification generation~\cite{wen2024enchanting, wu2023lemur}. 
However, these works do not tackle system-level modeling. 
By contrast, recent attempts at LLM-based formal model construction remain largely empirical and lack a systematic framework for the iterative procedure described above~\cite{hong2025effectiveness, capozucca2025ai}. 
Notably, the recently proposed PAT-Agent~\cite{zuo2025pat} is designed specifically for formal model construction, leveraging model checking. 
However, model checking alone does not suffice to prove the correctness of generated models. 
While model checking is effective at reporting counterexamples within a given bound, the absence of counterexamples only guarantees correctness up to that bound, and the cost of exploration grows rapidly as the bounds increase.
In contrast, theorem proving provides unbounded reasoning.
Proof obligations can be generated to ensure well-definedness of expressions, preservation of invariants, feasibility of events under their guards, and termination of the model \cite{abrial1996b}.
Consequently, correct-by-construction methods typically rely on a combination of model checking and theorem proving.
Similar to autoformalization, there has been growing interest in LLM-powered automated theorem provers~\cite{first2023baldur, lu2024proof, carrott2024coqpyt}. 
Yet most existing works focus on isolated tasks, such as theorem proving alone or model synthesis verified solely by model checking.
In practice, formal development requires models and proofs to evolve together.
When a theorem prover fails to discharge a proof obligation, the failure may stem from inadequate modeling or the inherent difficulty of automated reasoning.
Since neither model correctness nor prover completeness can be assumed, practical development must allow revisions to both the formal model and its proof artifacts.
To our knowledge, no existing automated approach formulates formal development as a joint state space over models and proof artifacts, enabling iterative coordination between model synthesis and proof-guided repair.

To bridge this gap, we propose \textbf{Event-B Agent}, a novel framework for end-to-end formal model construction powered by LLM. 
Given a requirement document in natural language, Event-B Agent constructs a formal model through three stages.
(1) \textbf{Refinement Stratey Planning.}
To allow incremental model design and reduce the complexity of both model synthesis and proof-guided repair, we adopt the LLM-proposed refinement strategy that distributes the requirements across multiple refinement steps.
With proof obligations, refinement guarantees that properties proven in earlier steps are preserved in later ones, thereby reducing the difficulty of synthesis and theorem proving.
(2) \textbf{Model Synthesis.}
For each refinement step determined by the refinement strategy, a formal model is synthesized from the assigned requirements and iteratively repaired to ensure well-formedness.
This process is crucial for effective model construction and achieving high requirement coverage.
(3) \textbf{Model \& Proof Repair.}
After synthesizing the model at a refinement step, it is verified and repaired as needed.
The model \& proof repair component discharges the proof obligations~\cite{abrial2010modeling} required for refinement soundness and requirement correctness.
This guarantees that properties established at earlier refinement levels are preserved, while newly introduced properties at the current level are verified.
The process is repeated until all proof obligations at the current refinement step are discharged, after which the synthesis-and-repair procedure continues for the next step.
The whole procedure stops when all the requirements have been addressed.

We evaluate Event-B Agent on 27 formal systems, each with an average of 182.41 proof obligations, assessing the consistency and correctness of the constructed models.
Since the existing works do not support joint evolution of models and proofs, we compare against the closest existing attempts with the same input and output setting, that is, the LLM-based autoformalization methods.
Across all evaluated metrics, Event-B Agent consistently outperforms the baselines, achieving a proof obligation discharge rate (PDR) of 97.86\%, and surpasses the baselines in requirement coverage (RC) and requirement fulfillment (RF) by 4.63\% and 18.01\%, respectively.
The ablation studies further demonstrate the effectiveness of the two key components of our framework: refinement strategy planning and model \& proof repair.
Moreover, these components enhance the performance of each other.
Refinement reduces the complexity of individual proofs, while the repair component establishes the correctness of refinements, ensuring that fulfilled requirements are preserved across refinement steps.
Additionally, Event-B Agent completes synthesis and repair for each system in an average of \SI{74.45}{minutes}, while discharging proof obligations takes an average of \SI{0.24}{minutes}.
In summary, this paper makes the following contributions:
\begin{enumerate}
  \item \textbf{Framework.} We introduce Event-B Agent, the first end-to-end framework for formal model synthesis and repair that supports refinement-based development with coordinated model construction and proof derivation, embodying the principle of correctness by construction.
  \item \textbf{Tool.} We implement a proof-of-concept tool to demonstrate the practicality of our approach and integrate it into Rodin~\cite{rodin}, an IDE for Event-B that provides formal verification support, including model checking, SMT solving, and theorem proving.
  \item \textbf{Evaluation.} We evaluate Event-B Agent on 27 formal systems against autoformalization baselines, including ablation studies to quantify component contributions, as well as efficiency and qualitative analyses. All code, datasets, and the interface are publicly available\footnote{\url{https://github.com/HongshuW/EventB_Agent}, \url{https://doi.org/10.5281/zenodo.19642103}}.
\end{enumerate}

\section{Motivation}

\subsection{Motivating Example}

\autoref{tab:example} lists the requirements of a formal system specifying an algorithm that searches for the minimum value of $f(j)$ over all valid $j$ in the domain of the function $f$. 
In this example, the requirements are divided into two categories: $EQP$ (equipment) and $FUN$ (function). 
The equipment requirements define the constants, sets, and axioms available to the system, while the function requirements specify its intended behavior. 
In this example, $EQP$--1 and $EQP$--2 introduce the constants $n$ and $f$.
$FUN$--1 states that upon termination, the algorithm must return an index $j \in dom(f)$ such that $f(j)$ is the minimum value in the range of $f$.
Requirements $FUN$--2 through $FUN$--6 describe the iterative searching process to achieve this goal.

\begin{table}[h!]
\centering
\footnotesize
\begin{tabular}{l p{0.85\linewidth}}
\toprule
\textbf{ID} & \textbf{Description} \\
\midrule
EQP-1 & Provides a finite natural number function $f$ defined on the domain $\{0..n-1\}$. \\
EQP-2 & Provides $n$ as the size of the domain of $f$. \\
FUN-1 & The final condition requires $j \in \text{dom}(f)$ and $f(j) = \min(\text{ran}(f))$. \\
FUN-2 & The system has a boolean variable \texttt{searching} to indicate whether the minimum has been fully identified. Two indices are maintained: $i$ as the scanning index and $j$ as the current candidate for the minimum. \\
FUN-3 & Initialization sets i = 1 and j = 0, with searching = true. \\
FUN-4 & While \texttt{searching} is true, if $i \neq n$ and $f(i) < f(j)$, update $j := i$ and increment $i$. \\
FUN-5 & While \texttt{searching} is true, if $i \neq n$ and $f(i) \geq f(j)$, keep $j$ unchanged and increment $i$. \\
FUN-6 & When $i = n$, \texttt{searching} becomes false, marking the completion of searching. \\
\bottomrule
\end{tabular}
\caption{Equipment (EQP) and Function (FUN) requirements of minimum-searching model.}
\label{tab:example}
\end{table}

\begin{figure}[t]
\includegraphics[width=\linewidth]{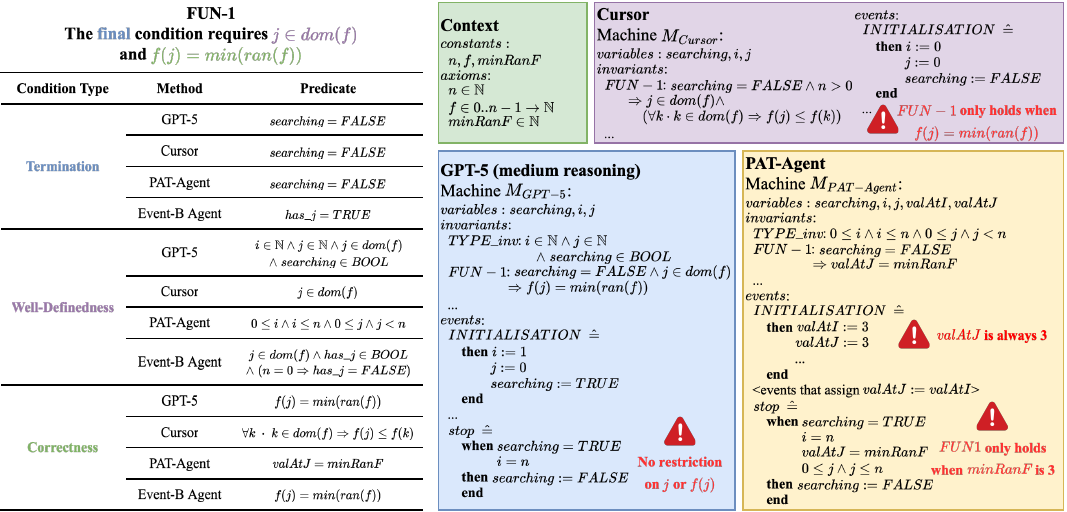}
\caption{Simplified Event-B models for the motivating example generated by GPT-5, Cursor, and PAT-Agent. }
\label{fig:example_new}
\centering
\end{figure}

\begin{figure}[t]
\includegraphics[width=0.85\linewidth]{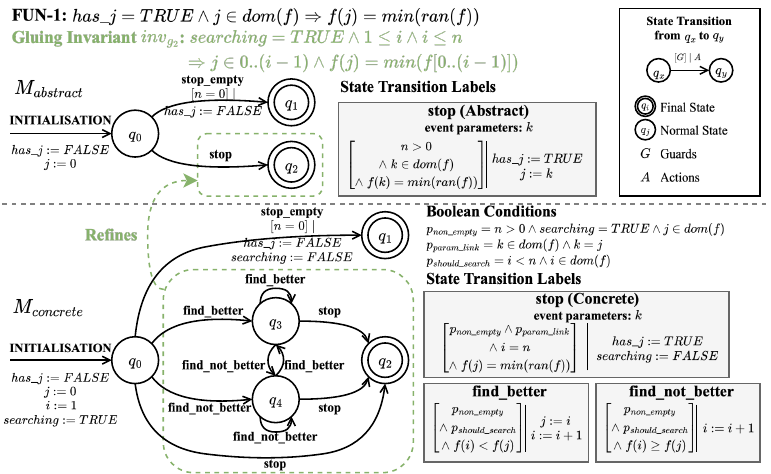}
\caption{The abstract and concrete formal models constructed by Event-B Agent for the motivating example.}
\label{fig:running_example}
\centering
\end{figure}

As discussed in Section~\ref{sec:intro}, we select a set of existing LLM-based autoformalization approaches, namely GPT-5 with medium reasoning, Cursor, and PAT-Agent, and adapt them for the task of formal model construction.
Their performances are illustrated in \autoref{fig:example_new}, while the formal models constructed by Event-B Agent are visualized in \autoref{fig:running_example}.
The ``Context'' block in \autoref{fig:example_new} represents the shared context across the models generated by various methods, corresponding to the EQP requirements.
The table on the left of the figure shows how each method formalizes requirement $FUN$--1 to express the algorithm termination condition, well-definedness, and correctness condition.
In model $M_{Cursor}$ generated by Cursor, the $INITIALISATION$ event sets $j:=0$ and $searching:=FALSE$, violating $FUN$--1 unless $f(0)=min(ran(f))$.
In contrast, the remaining baselines do not mistakenly impose termination constraints during initialization, but exhibit other limitations.
GPT-5 produces model $M_{GPT-5}$ with $FUN$--1: $searching=FALSE \; \land \; j \in dom(f) \Rightarrow f(j)=min(ran(f))$, where the termination condition is not required to hold in the initial state.
However, in the $stop$ event, the guards do not restrict $j$ or $f(j)$, allowing the algorithm to terminate with an arbitrary $j$ and violate $FUN$--1.
PAT-Agent, a framework specialized for formal model synthesis, produces a stronger model $M_{PAT-Agent}$ in which the guards of $stop$ enforce $valAtJ=minRanF$, thereby satisfying $FUN$--1.
Nonetheless, this model contains subtle flaws.
In $INITIALISATION$, both $valAtI$ and $valAtJ$ are set to $3$, and subsequent events only update $valAtJ:=valAtI$ without modifying $valAtI$. 
As a result, the guard of $stop$ implies that the constant $minRanF$ must always be $3$. 
From the perspective of model checking, this system is deemed correct, but the correctness does not generalize beyond the single case $minRanF=3$.

The models generated by Event-B Agent, $M_{abstract}$ and $M_{concrete}$, resolve this issue through refinement.
The models are shown in \autoref{fig:running_example}, where events are represented by state transitions.
We use $[G]|A$ to denote an event with guards $G$ and actions $A$.
Since $FUN$--1 depends only on $j$ and $f$, the abstract model omits auxiliary variables such as $i$ and $searching$, and focuses solely on proving $FUN$--1.
$FUN$--1 is vacuously true for $INITIALISATION$ and $stop\_empty$ events.
For the $stop$ event, $FUN$--1 is preserved because the event’s assignments ensure $has\_j = TRUE \land j \in dom(f)$ in the post-state, and the guard enforces $f(j) = min(ran(f))$.
By construction of the refinement, the refined events $INITIALISATION$, $stop$, and $stop\_empty$, strengthen the guards of their abstract counterparts and simulate the abstract assignments, thereby preserving the invariant in $M_{\text{concrete}}$.

This example highlights three key insights motivating the design of Event-B Agent.
First, refinement reduces modeling and proving complexity by abstracting away irrelevant details. 
Second, when models are inaccurate, a systematic repair framework is required, e.g., updating the actions of $INITIALISATION$ in $M_{Cursor}$ and the guards of $stop$ in $M_{GPT-5}$.
Finally, model checking alone is insufficient to guarantee correctness, and proof information must be exploited in model synthesis and repair.
Accordingly, we propose Event-B Agent, which addresses the formal model synthesis and repair problem by incorporating refinement, a novel repair component, and proof information.

\subsection{Problem Statement} \label{sec:problem_statement}

Let $M$ denote the set of formulas extracted from a formal model, $\pi$ denote a proof artifact witnessing that a proof obligation (PO) $\varphi$ is derivable from $M$ under a sound proof system $\mathcal{R}$.
Then $M$ is said to discharge $\varphi$ if and only if there exists $\pi$ that witnesses $M \vdash_{\mathcal{R}} \varphi$.
We abstract formal development as a process over model construction and proof derivation, where both may be updated independently or simultaneously within each step.
Formally,
\[
(M^0, \pi_0) \rightsquigarrow (M^1, \pi_1) \rightsquigarrow ... \rightsquigarrow (M^n, \pi_n)
\text{ s.t. }
\forall \; t \in \{0,\dots,n\}, \pi_t \text{ is a valid proof for } M^t \vdash_{\mathcal{R}} \varphi_t,
\]
where $M^0 = \varnothing$, $\pi_0 = \epsilon$, $n$ is the number of POs, and the step $(M^0, \pi_0) \rightsquigarrow (M^1, \pi_1)$ synthesizes the initial model $M^1$ given natural language requirements and discharges $\pi_1$.
Each pair $(M^t, \pi_t)$ represents an intermediate model $M^t$ with a proof $\pi_t$ discharged at step $t$, allowing updates to both $M^t$ and $\pi_t$, while ensuring soundness at each step through $\pi_t$ under the sound proof system $\mathcal{R}$.

We position our work relative to two lines of research.
LLM proof assistants assume a fixed model $M$ and search for valid $\pi$, but cannot construct $M^0 \rightsquigarrow M^1$ from natural language requirements or revise $M^t$ during development.
In contrast, LLM-based autoformalization methods iteratively construct models $M^0 \rightsquigarrow \cdots \rightsquigarrow M^k$ without proof-based verification or repair, relying primarily on bounded model checking with limited correctness guarantees.
Neither line of work formulates formal development as a coordinated process of model construction and proof derivation.

Furthermore, refinement should be supported to incrementally develop the models.
Let $M_i^t$ denote the $t$-th model at refinement level $i$, where $i=\{1,\dots,m\}$.
For each $i < m$, the last model at level $i$ is refined by the first model at level $i+1$, with refinement soundness ensured by POs.
Eventually, $M_m^n$ is produced as the final model.
Existing LLM-based autoformalization does not support refinement in this sense, and LLM proof assistants operate over fixed models without refinement.
\section{Preliminary}\label{sec:preliminary}

\begin{figure}[t]
\centering
\footnotesize
\setlength{\jot}{0pt}

\begin{minipage}[t]{0.4\columnwidth}
\[
\begin{alignedat}{2}
i &\;::=\; \textit{literal}; &\quad& \text{Identifier} \\
il &\;::=\; \{i\}; & & \text{List of identifiers} \\
pred_{il} &\;::=\; predicate; & & \text{Predicate such that } \\
 & & & \quad FV(pred_{il}) \subseteq il \\
M &\;::=\; {Context}, & & \text{Formal model} \\
 & \quad {Machine}; & & \\
 & & & \\
Context &\;::=\; i, & & \text{Context identifier} \\
 & \quad [\text{``extends''},i], & & \text{Extended context} \\
 & \quad \text{``sets''},s, & & s ::= il \\
 & \quad \text{``constants''},c, & & c ::= il \\
 & \quad \text{``axioms''},A, & & A ::= \{pred_{s \cup c}\} \\
 & \quad \text{``theorems''},T; & & T ::= \{pred_{s \cup c}\} 
\end{alignedat}
\]
\end{minipage}
\hfill
\begin{minipage}[t]{0.57\columnwidth}
\[
\begin{alignedat}{2}
Machine &\;::=\; i, &\quad& \text{Machine identifier} \\
 & \quad [\text{``refines''},i], & & \text{Refined machine} \\
 & \quad [\text{``sees''},i], & & \text{Seen context} \\
 & \quad \text{``variables''},v, & & v ::= il\\
 & \quad \text{``invariants''},I, & & I ::= \{pred_{s \cup c \cup v}\} \\
 & \quad \text{``variants''},N, & & N::= \{pred_{s \cup c \cup v}\} \\
 & \quad \text{``theorems''},T, & & T ::= \{pred_{s \cup c \cup v}\} \\
 & \quad \text{``events''},\mathcal{E}; & & \mathcal{E} ::= \{Event\} \\
 & & & \\
Event &\;::=\; i, & & \text{Event identifier} \\
 & \quad \text{``any''},x, & & \text{Event parameters, } x ::= il \\
 & \quad \text{``where''},G, & & \text{Event guards, } G ::= \{pred_{s \cup c \cup v \cup x}\} \\
 & \quad \text{``then''},v \;:|\; BA; & & \text{Actions, } BA ::= \{pred_{s \cup c \cup v \cup x \cup v'}\} \\
\end{alignedat}
\]
\end{minipage}

\caption{The grammar of Event-B, where \(FV(pred)\) is the set of free variables in \(pred\).}
\label{fig:grammar}
\end{figure}

\textbf{The Event-B Notation.}
\autoref{fig:grammar} summarizes the grammar of Event-B \cite{abrial2010modeling}.
An Event-B model $M$ is a discrete transition system grounded in set theory and first-order logic, where all types are represented as sets.
For instance, $STRING$ denotes the set of all strings, and $A \rightarrow B$ the set of total functions with domain $A$ and range $B$.

A model $M$ consists of \emph{contexts} and \emph{machines}.
Let $il$ be a list of identifiers. Sets $s$, constants $c$, variables before and after an event $v$ and $v'$, and event parameters $x$ are defined over $il$ and are disjoined.
A predicate $pred$ with $FV(pred) = il$ is denoted by $pred_{il}$, where $FV(pred)$ is the set of free variables in $pred$.
A list of predicates using only variables in $il$ is denoted by $\{pred_{il}\}$, which can be used to express components like axioms, $A ::= \{pred_{s \cup c}\}$.

\noindent
\textbf{Correctness by Construction.}
Let $REQ_M$ denote the requirements of formal model $M$, and $PO_M$ the set of its proof obligations.
Each $req \in REQ_M$ is associated with a predicate $Covered(M, req) \in \{\mathsf{true}, \mathsf{false}\}$, which holds iff $req$ is covered in $M$ as an invariant, variant, axiom, or event.
Similarly, each $po \in PO_M$ has a predicate $Discharged(M,po) \in \{\mathsf{true}, \mathsf{false}\}$, which holds iff $po$ is discharged.
We define a predicate $Correct(M)$ to indicate that $M$ is \emph{correct by construction}, namely:
{\footnotesize\begin{equation}
Correct(M) \;\triangleq\;
\Bigl(\bigwedge_{req \in REQ_M} Covered(M, req)\Bigr)
\;\wedge\;
\Bigl(\bigwedge_{po \in PO_M} Discharged(M,po)\Bigr)
\label{eq:correctness}
\end{equation}}

\section{Methodology} \label{sec:method}

\subsection{Overview}

\begin{figure}[t]
\includegraphics[width=0.9\linewidth]{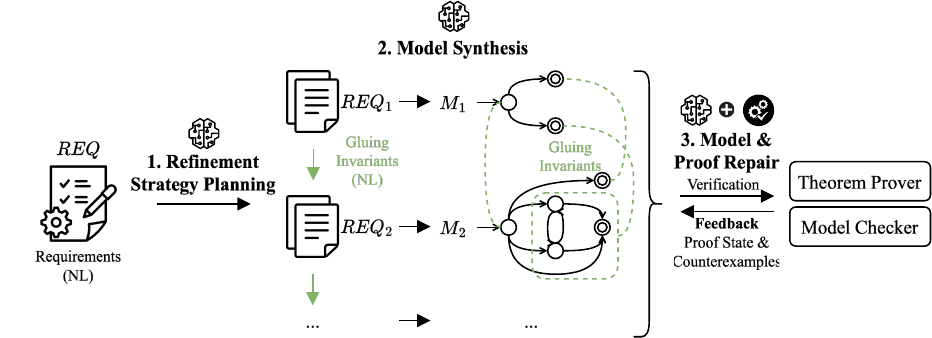}
\caption{Overview of Event-B Agent. From natural-language requirements ($REQ$), the system plans a refinement strategy to guide incremental model synthesis, with verification and repair at each refinement level.}
\label{fig:Overview}
\centering
\end{figure}

In this work, we propose Event-B Agent, an LLM-powered end-to-end framework for formal model synthesis and repair.
\autoref{fig:Overview} shows the overall workflow of Event-B Agent, the model \& proof repair component is elaborated in \autoref{fig:modelrepair_v2}, and \autoref{fig:running_example} illustrates the minimum-searching model generated by Event-B Agent.
System description in natural language is provided to Event-B Agent to generate a refinement strategy, which distributes requirements to various refinement steps.
In this process, gluing invariants between two refinement steps are summarized in natural language.
During model synthesis, the distributed requirements and gluing invariants are formalized into a model.
The correctness of the model is then verified through model checking and theorem proving, and repairs are applied accordingly.
This procedure is repeated for each refinement step until all proof obligations are discharged or the trial limit is reached.

Event-B Agent adopts a neurosymbolic design, where semantic tasks such as refinement planning, model synthesis, and repair are delegated to specialized LLMs, while deterministic components handle the rest.
Model checkers, SMT solvers, and theorem provers verify synthesized models, pattern matching identifies candidate repair rules from the proof state, and atomic repair functions apply updates to models and proofs.
This separation combines the flexibility of LLMs with the reliability of symbolic reasoning, achieving both semantic versatility and soundness.

\subsection{Refinement Strategy Planning}

Automated reasoning over realistic software systems is challenging due to the rapid growth of the search space as model complexity increases.
Refinement offers a complementary approach by establishing properties in simpler abstract models, where proofs are easier to discharge due to abstraction, and preserving them in more detailed concrete models through refinement proof obligations.
This ensures that proof effort scales with model complexity while preserving correctness.

On this basis, Event-B Agent utilizes LLMs to automatically plan refinement strategies from natural language requirements. 
Refinement introduces additional detail in a disciplined manner, guided by gluing invariants that relate abstract and concrete models to maintain consistency.

\subsubsection{Refinement} \label{sec:refinement}

Having introduced the role of refinement, we now formalize this notion.
Let $M_A$ denote an abstract model and $M_C$ a concrete refinement of $M_A$.
Let $REQ_{M}$ denote the set of requirements encoded in model $M$.
Refinement preserves requirements of the abstract model $REQ_{M_A}$ while introducing additional ones $REQ_{\mathit{diff}}$, i.e.,
\(
REQ_{M_A} \subset REQ_{M_C}
\wedge
REQ_{\mathit{diff}} \;\triangleq\; REQ_{M_C} \setminus REQ_{M_A}
\).

The preservation of proven properties is guaranteed through gluing invariants and the corresponding refinement POs.
For instance, the refined events are required to have stronger guards than the abstract events, and their actions must simulate the abstract counterparts.
With these guarding POs, Event-B Agent ensures refinement correctness, as illustrated in \autoref{fig:running_example}.

\subsubsection{Gluing Invariant} \label{sec:gluinginv}

A gluing invariant is a predicate over both abstract variables and concrete variables, specifying how states in the concrete model correspond to states in the abstract model.
\autoref{fig:running_example} includes a gluing invariant between $M_{abstract}$ and $M_{concrete}$ generated by Event-B Agent.
$M_{concrete}$ introduces a variable $i$ to represent the current index during search, and a boolean variable $searching$ to represent whether the searching procedure is ongoing.
The gluing invariant relates the abstract variable $j$ to the two concrete variables $i$ and $searching$, and specifies that during searching, $f(j)$ is always the minimum value among the explored domain $0..(i-1)$.
The model synthesis step later formalizes it into $searching=TRUE \land 1\leq i \land i \leq n \Rightarrow j \in 0..(i-1) \land f(j) = min(f[0..(i-1)])$.

Gluing invariants are added to the model $M$ and play a critical role in discharging refinement POs.
In Event-B Agent, candidate gluing invariants are first proposed by the refinement strategy planning LLM and then formalized during model synthesis.
Since LLM-based generation offers no correctness guarantee, gluing invariants are validated in two steps: (1) counterexample checking and contradiction detection with the model checker, and (2) attempting proofs relevant to the gluing invariants before other proofs.
Only after no counterexamples are found and the proofs succeed, the gluing invariants will be accepted into $M$ for subsequent reasoning.

\subsubsection{Refinement Strategy}

Refinement proceeds inductively, distributes requirements across successive steps, and preserves established properties.
The first stage of Event-B Agent is to plan a refinement strategy that decides how requirements are allocated across steps and how successive models are linked by gluing invariants.
Once the strategy is set, the model synthesis and repair components construct the models, verify the requirements, and apply corrections if needed.

Concretely, the refinement strategy planning step of Event-B Agent partitions $REQ$ into $n$ disjoint subsets $REQ_{M_i}$ such that $REQ=\bigcup_{i=1}^{n} REQ_{M_i}$.
$REQ_{M_1}$ includes the requirements for the initial abstract model, $REQ_{M_2}$ contains the additional requirements for the second model, etc.
Between steps $i-1$ and $i$, the refinement strategy planning LLM also proposes a set of gluing invariants $\mathcal{I}_{g_i}$, which are formalized and added into $M_i$ during model synthesis.
Thus, the additional requirements for step $i$ are $REQ_{M_i} \cup \mathcal{I}_{g_i}$, where $REQ_{M_i}$ specifies system requirements and $\mathcal{I}_{g_i}$ aims to preserve all previously satisfied requirements $\bigcup_{j=1}^{i-1} REQ_{M_j} \cup \mathcal{I}_{g_j}$, where $\mathcal{I}_{g_1}=\varnothing$.

In the minimum-searching example, Event-B Agent creates a refinement strategy of 2 steps, $REQ_{M_1} = \{FUN\text{--1}\}$ and $REQ_{M_2} = \{FUN\text{--2}, FUN\text{--3}, FUN\text{--4}, FUN\text{--5}, FUN\text{--6}\}$.
Additionally, as shown in \autoref{fig:running_example}, a set of gluing invariants $\mathcal{I}_{g_2} = \{inv_{g_2}\}$ is generated to capture the relation between refinement steps 1 and 2.
At this stage, the requirements in $REQ_{M_1}$, $REQ_{M_2}$, and $\mathcal{I}_{g_2}$ are in natural language form and will be formalized during the model synthesis step.

Formally, the correctness of a refined model $M_i$ with respect to all requirements up to step $i$ is defined inductively as:
{\footnotesize \begin{equation}
    Correct(M_i) \;\triangleq\; 
    \Bigl(\bigwedge_{j=1}^{i-1} Correct(M_j)\Bigr) 
    \land \Bigl(\bigwedge_{req \; \in \;REQ_{M_i} \cup \mathcal{I}_{g_i}} Covered(M_i, req)\Bigr) 
    \land \Bigl(\bigwedge_{po \in PO_{M_i}} Discharged(M_i, po)\Bigr)
\label{eq:refinement_correctness}
\end{equation}}

This planning stage provides the blueprint for synthesis and repair: each step has well-defined requirements and invariants, while correctness is ensured and preserved through proof obligations.

\subsection{Model Synthesis} \label{sec:modelsynthesis}

For each refinement step $i$ with requirements $REQ_{M_i} \cup \mathcal{I}_{g_i}$, Event-B Agent synthesizes a formal model that captures the specified requirements.
The model must be syntactically correct and free of compilation errors so that formal verification tools can be applied effectively.

\noindent
\textbf{Schema-Guided Formalization.} 
Simple prompt-based generation frequently produces ill-formed code, including undeclared variables, malformed invariants, and syntactic errors that are not parsable, particularly in the synthesis of formal specifications where training data is sparse.
To address this, we design a JSON schema that encodes the grammar of Event-B shown in \autoref{fig:grammar}.
The schema encodes structural constraints, for example, every machine must contain at least one event, and may include variables, invariants, and variants.
Because the schema is language-agnostic, it can be readily adapted to other formal specification languages, making our approach generalizable.

\noindent
\textbf{Synthesis and Repair Loop for Well-Formedness.}
The model synthesis LLM first generates a candidate model according to the JSON schema.
While the schema eliminates most syntax errors, it cannot enforce typing constraints, so issues such as undeclared variables, type mismatches, and invariants referencing undefined sets may still arise.
The candidate model is parsed and compiled into Event-B code.
Compilation errors are fed back to the LLM, together with the current model, for iterative repair to ensure well-formedness.
This synthesis–repair loop ensures that models passed to verification tools are syntactically valid and type-correct, providing a reliable foundation for subsequent refinement and semantic repair.

\noindent \textbf{Ensuring Refinement Soundness.}
While synthesizing a refined model, the requirements at the current refinement level and the previous abstract model are provided to the LLM as context.
The refinement relation is preserved through refinement POs after generating the refined model.
In our experiments, the refinement PO discharge rate is collected to evaluate the refinement correctness.

\subsection{Model \& Proof Repair}
\label{sec:repair}
\begin{figure}[t]
\includegraphics[width=\linewidth]{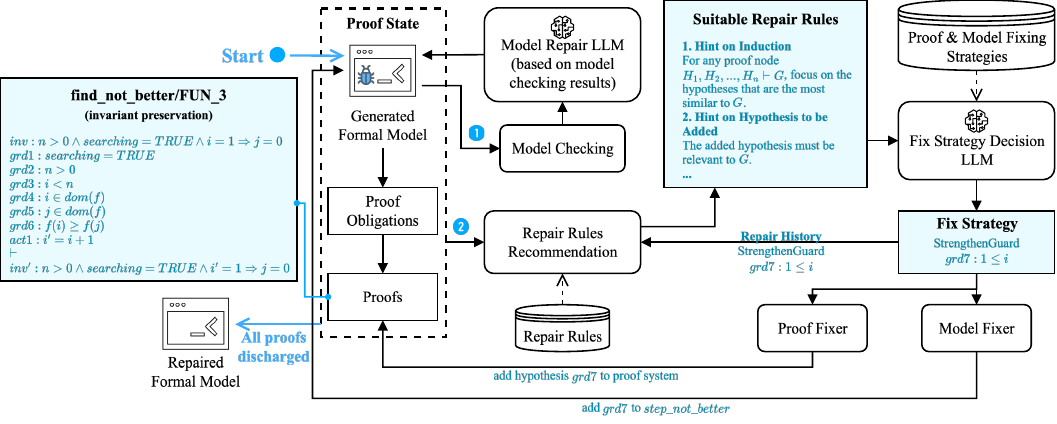}
\caption{Workflow of the model \& proof repair component. The model is \textcircled{1} checked by a model checker (bounded) and repaired, then \textcircled{2} repaired guided by unbounded theorem proving.
}
\label{fig:modelrepair_v2}
\centering
\end{figure}

A key contribution of our work is the model \& proof repair component, which iteratively verifies models, diagnoses inaccuracies, and updates both models and proofs.
An overview of this component is shown in \autoref{fig:modelrepair_v2}.

\subsubsection{Formal Verification and Repair}

Whenever a new model is synthesized, it is first verified by a model checker to detect violations of invariants, including both requirement and gluing invariants (Step \textcircled{1} in \autoref{fig:modelrepair_v2}).
The model repair LLM suggests suitable bounds, within which the model checker searches for states that violate invariants $\mathcal{I}$.
If a counterexample trace is found, it indicates an invariant violation. This trace is then passed to the model repair LLM as feedback for correcting the model to resolve the violation within the given bounds.

While model checking is effective for detecting invariant violations within the given bounds, it cannot guarantee full correctness.
This is because bounded analysis may miss unreachable states or behaviors outside the explored bounds.
To cover all executions, theorem proving is applied in addition.
After proof obligations are generated, automated provers and constraint solvers attempt to discharge them under the relevant hypotheses extracted from the model.
When proofs fail, the cause may range from modeling inaccuracies or limitations of the automated reasoning process.
To address this, the model \& proof repair component analyzes the proof state and applies targeted fix strategies to either the model or its proofs (Step \textcircled{2} in \autoref{fig:modelrepair_v2}).

\begin{table}[h!]

\footnotesize
\centering

\begin{tabular}{p{0\linewidth} p{0.13\linewidth} p{0.23\linewidth} p{0.53\linewidth}}
\toprule
\textbf{} & \textbf{Category} & \textbf{Condition} & \textbf{Rules} \\ 

\midrule
1 & \begin{tabular}[c]{@{}l@{}}
Contradictory \\
Goal
\end{tabular} & 
\begin{tabular}[c]{@{}l@{}}
Goal is $\bot$
\end{tabular} & 
\begin{tabular}[c]{@{}l@{}}
1. PO is about an invariant $\rightarrow$ update the invariant to eliminate \\
contradictions \\
2. PO is about an event $\rightarrow$ add or update actions or guards to \\
eliminate contradictions
\end{tabular} \\

\midrule
2 & \begin{tabular}[c]{@{}l@{}}
True by \\
Definition 
\end{tabular} & 
\begin{tabular}[c]{@{}l@{}}
Goal matches definitional \\
pattern, 
e.g. $x=y$, $x \in S$ etc.
\end{tabular} & 
\begin{tabular}[c]{@{}l@{}}
1. Goal holds by definition $\rightarrow$ apply the corresponding tactics \\
2. (a) Rule 1 doesn't update $\mathcal{P}$, or (b) Goal is false $\rightarrow$ Similar rules\\
as \textbf{Category 1}
\end{tabular} \\

\midrule
3 & Existential Goal & 
\begin{tabular}[c]{@{}l@{}}
Goal is an $\exists$ predicate, \\
e.g. $\exists x \cdot <predicate>$
\end{tabular} & 
\begin{tabular}[c]{@{}l@{}}
1. $x$ is a variable $\rightarrow$ add or update actions to define $x$ \\
2. $x$ is a constant $\rightarrow$ add hypothesis as an axiom in the context to \\
establish its existence
\end{tabular} \\

\midrule
4 & 
\begin{tabular}[c]{@{}l@{}}
Equality PO
\end{tabular} & 
\begin{tabular}[c]{@{}l@{}}
PO type is Equality, e.g. goal \\
$G$ is $x=y$, where $x$, $y$ are \\
from machines at different \\
refinement steps
\end{tabular} & 
\begin{tabular}[c]{@{}l@{}}
1. If there is an indirect relation $H$ (e.g. $f(x)=f(y)$), add axiom \\
of the form $H \Rightarrow G$ if suitable.
\end{tabular} \\

\midrule
5 & \begin{tabular}[c]{@{}l@{}}
Well-\\
Definedness 
\end{tabular} & 
\begin{tabular}[c]{@{}l@{}}
PO type is Well-Definedness \\
e.g. ensure $y \neq 0$ for $x/y$, \\
ensure $x \in dom(f)$ for $f(x)$
\end{tabular} & 
\begin{tabular}[c]{@{}l@{}}
1. PO is about guard $\rightarrow$ strengthen the guard based on the \\
definition of well-definedness \\
2. PO is about invariant $\rightarrow$ strengthen the invariant based on the \\
definition of well-definedness
\end{tabular} \\

\midrule
6 & 
\begin{tabular}[c]{@{}l@{}}
Quantified \\
Invariant \\
Preservation
\end{tabular} & 
\begin{tabular}[c]{@{}l@{}}
PO is Invariant Preservation \\
about a quantified invariant
\end{tabular} & 
\begin{tabular}[c]{@{}l@{}}
1. Universal invariant $\rightarrow$ Find quantified hypothesis \\
$\forall x \cdot H(x) \Rightarrow G(x)$, where $G(x)$ is similar to proof goal $G(x')$. \\
Instantiate the hypothesis with $x'$. \\
2. Existential invariant $\exists x \cdot P(x)$ $\rightarrow$ Similar 
rules as \textbf{Category 3}
\end{tabular} \\

\midrule
7 & 
\begin{tabular}[c]{@{}l@{}}
Uninstantiated \\
Hypothesis
\end{tabular} & 
\begin{tabular}[c]{@{}l@{}}
Previous repairs add \\
quantified hypotheses that \\
are not yet instantiated
\end{tabular} & 
\begin{tabular}[c]{@{}l@{}}
1. Quantified hypothesis added but not 
instantiated $\rightarrow$ suggest \\
suitable instantiation values
\end{tabular} \\

\bottomrule
\end{tabular}

\caption{Categories of proof states and the corresponding repair rules supported by Event-B Agent.}
\label{tab:rule-categories}

\end{table}

\subsubsection{Repair Rules Recommendation} \label{sec:proofrules}
To improve the reasoning capability of the fix strategy decision LLM, we design a novel repair rules recommendation component that suggests suitable rules for repairing the given model $M$ and a proof $\pi$ about the model, and provides them as guidance to the LLM.
We observe that characteristic patterns emerge across model $M$, proof $\pi$, the proof obligation type, and their corresponding historical repairs $R(M, \pi)$.

To illustrate how repair rules are recommended, consider the case of universally quantified hypotheses. 
In interactive theorem proving, such hypotheses become useful only after instantiation with concrete terms.
Although SMT solvers perform heuristic instantiation, this process is incomplete and often fails in Event-B proofs.
Thus, after adding a universal predicate as a lemma in $\pi$, the next step is typically to instantiate it. 
If the repair history $R(M, \pi)$ indicates that a universal lemma has been added but no instantiations appear, and the child nodes in the proof tree remain open, the following rule is recommended:
\textit{``If a universal lemma has been added to a node in the proof tree, but its child nodes remain unclosed, suggest instantiation values for the lemma''}.  
In this example, the first action modifies $M$ by adding the universal lemma as a reusable axiom in the context, while the second action instantiates it in $\pi$ if SMT solvers do not already discharge the proof.

Beyond this example, our system supports a total of seven categories of rules, summarized in \autoref{tab:rule-categories}. 
The ``Condition'' column specifies how each category is determined from the current proof state, including information from the model $M$, proof $\pi$, repair history $R(M, \pi)$, and proof obligation type.
Within each category, multiple concrete repairs may apply, and these sometimes cannot be determined by pattern matching. 
Instead, the fix strategy decision LLM analyzes the scenario, considers the recommended rules, and attempts repairs. 
The ``Rules'' column lists a non-exhaustive set of simplified rules under various scenarios.
These categories and rules were derived empirically based on our experience in manually discharging proofs.
While not exhaustive, they capture recurring patterns observed across a wide range of models.
If the current proof state does not fall into one of the known categories, the LLM falls back on a set of general default rules.

\autoref{fig:modelrepair_v2} shows an example of model-proof repair with the default rules.
$FUN$--3 in \autoref{tab:example} specifies a property about initial values of variables, formalized as $inv: \; n>0 \land searching=TRUE \land i=1 \Rightarrow j=0$.
The PO $find\_not\_better/FUN\_3$ fails because an event instance with pre-state $i=0$ and post-state $i'=1$ is admitted, yielding a post-state invariant $inv'$ that requires $j=0$, which cannot be derived from the proof context.
However, such a state is unreachable in the actual execution of the algorithm, where $i \geq 1$ is maintained (cf. \autoref{fig:running_example}).
As the proof state does not match any rule category in \autoref{tab:rule-categories}, the repair rules recommendation component falls back to default rules, whose simplified forms are shown as ``Suitable Repair Rules'' in \autoref{fig:modelrepair_v2}.
These rules guide the later steps to focus on the inductive relationship between $inv$ and $inv'$ (Rule 1), and add relevant hypotheses to $inv'$ when needed (Rule 2).

\subsubsection{Fix Strategy Selection and Execution}\label{sec:fix_selection}

The repair rules introduced in Section~\ref{sec:proofrules} are used to guide the fix strategy decision LLM, which may hallucinate during repair.
To mitigate this issue, Event-B Agent restricts modifications to a library of atomic repair functions (e.g., strengthen an invariant) that correspond to a set of fix strategies to update models and proofs.
The functions are described as ``atomic'' because they represent the smallest modification units.
After the repair rules are obtained, the LLM selects an atomic function and proposes the corresponding parameters, and the proof state is updated after executing the function.
This iterative process continues until the proof is discharged or the trial limit is reached.
The commonly used fix strategies are summarized in \autoref{fig:distribution}, which are grouped into 4 categories:
\begin{itemize}
    \item \textbf{Model modification strategies}: This group of strategies corresponds to atomic functions such as adding or updating invariants, strengthening guards, and modifying actions;
    \item \textbf{Proof modification strategies}: These strategies correspond to atomic functions such as instantiating quantified hypotheses and unfolding definitional equalities;
    \item \textbf{Joint model–proof modification strategies}: Operations that simultaneously modify the model and the selected proof, e.g., introducing a hypothesis as a context axiom while injecting it into the proof context, thereby repairing both in tandem;
    \item \textbf{Information Retrieval strategies}: Operations such as invoking the model checker to obtain guidance for subsequent repair steps, without directly modifying the model or proof.
\end{itemize}

To summarize, the fix strategy decision LLM performs two complementary tasks: 
(1) selecting the appropriate function based on the current proof state and recommended repair rules; and 
(2) proposing the values of function parameters, which specify the repair contents.
For instance, in \autoref{fig:modelrepair_v2}, the selected function is $strengthenGuard$, with parameters $guard\_label$ and $guard\_content$.
The repair adds $grd7:1 \leq i$, inferred from the inductive relationship between $inv$ and $inv'$.

\subsubsection{Soundness of Repair.}
In our method, repairs aimed at discharging POs are generated jointly by the repair rule recommendation component and the fix strategy decision LLM.
Based on pattern matching over the proof state, the former retrieves repair rules to construct prompts for the latter.
A sequence of repairs is accepted only if their combined effect discharges the target PO.
Once discharged, SMT solvers and theorem provers guarantee its validity.
After each model modification, all proofs are replayed to ensure that any invalidated proofs are not considered successful.
As a result, soundness is ensured by the verification pipeline rather than the repair rules.
The rules provide heuristic guidance by suggesting common proof actions, and our ablation study shows clear empirical gains from this repair guidance mechanism.
Importantly, they do not alter the underlying modeling or proof mechanisms and therefore do not affect soundness.

\section{Evaluation}

Event-B Agent is evaluated through the following research questions:
\begin{itemize}
    \item \textbf{RQ1.} Is Event-B Agent effective in constructing consistent and correct models? How does it compare against existing approaches attempting this problem?
    \item \textbf{RQ2.} How does each component of Event-B Agent contribute to its overall performance?
    \item \textbf{RQ3.} How efficient is Event-B Agent in the task of formal model synthesis and repair?
    \item \textbf{RQ4}. How does Event-B Agent construct and repair models across different refinement steps, and what atomic repair functions does it employ?
\end{itemize}

\subsection{Experimental Setting} \label{sec:evalsetting}
\textbf{Tool.}
We developed a proof-of-concept implementation of Event-B Agent integrated into the Rodin IDE \cite{rodin}.
The structured JSON schema described in Section~\ref{sec:modelsynthesis} enables the generated Event-B specification to be parsed and incorporated into the IDE, which natively supports structured Event-B code.
In our experiments, model checking was performed by ProB \cite{ProB}, a model checker for Event-B that checks deadlock-freeness, liveliness, consistency of axioms, and invariants preservation.

\noindent\textbf{Backbone LLM.}
We use GPT-5 (medium reasoning configuration, 2025-08-07 version)~\cite{openai_gpt5} as the backbone LLM in our experiments to demonstrate that formal model synthesis and repair remain challenging even for one of the most advanced LLMs.

\noindent
\textbf{Baselines.}
To the best of our knowledge, no prior work formulates formal development as a coordinated process of model construction and proof derivation to achieve correct-by-construction development.
For empirical evaluation, we therefore consider the closest existing approaches that share the same input–output setting.
For comparison, the outputs must support proof obligation generation and theorem proving.
A common specification language with such support is therefore required.
We chose Event-B because it provides unbounded proving and model checking, enabling the comparison and allowing the model checking based approaches to operate.
Specifically, we evaluate against three representative baselines below:

\begin{enumerate}
    \item \textbf{LLM Model Synthesis with Automated Provers.} A straightforward baseline is to generate specifications with an LLM and discharge the resulting proof obligations using automated provers. Within Rodin, this includes built-in engines such as PP (Predicator Prover) and integrated SMT solvers (CVC4 \cite{barrett2011cvc4}, Z3 \cite{de2008z3}, etc.).
    \item \textbf{Adapted General Purpose Coding Agent (Cursor).} We evaluate Cursor~\cite{CursorAI}, a commercial coding LLM agent that integrates features such as file editing, codebase search, and terminal execution. In our setup, web search is disabled, while other features remain enabled. Cursor is guided by high-level task instructions and explicit commands to parse Event-B specifications (JSON) into a formal model and run the model checker.
    \item \textbf{Adapted PAT-Agent.} To evaluate existing formal agents in our problem setting, we adapt PAT-Agent~\cite{zuo2025pat}, originally designed for autoformalization in PAT \cite{sun2009pat}. PAT and Event-B are both event/state-based specification languages, differing mainly in syntax and prover support. Our adaptation generates Event-B models by mapping context and machine constructs to PAT’s constants, variables, guarded actions, and processes; rewriting syntax documentation and examples in Event-B; replacing the PAT model checker with ProB for verification feedback; and omitting proof obligations, as PAT-Agent focuses solely on model checking. This adapted version serves as a baseline to contrast with Event-B Agent’s dual approach, integrating both model checking and theorem proving.
\end{enumerate}
After model synthesis by Cursor and PAT-Agent, we apply the automated provers from baseline~(1) to their outputs, ensuring all baselines are evaluated under comparable conditions.

\noindent \textbf{Dataset.}
We collected a dataset with 27 formal systems, including classic examples and algorithms collected by Jean-Raymond Abrial (the designer of the B Method and Event-B) and widely regarded as representative Event-B developments~\cite{abrial2010modeling}, together with real-world systems~\cite{riviere2025extending, riviere2023formalising}.
For the dataset on classic algorithms and real-world systems, we manually construct the requirement document based on the descriptions of the systems.
We partition the dataset into three subsets, ``Simple'', ``Medium'', and ``Complex'', based on the number of requirements (3–8, 9–13, and 14–24, respectively), with nine systems in each partition.
Although requirement count is not the only possible criterion, it is available prior to model construction and correlates with modeling complexity, as reflected by the increasing average number of POs in Event-B models generated by our method (89.22, 173.7, and 284.3).
Other potential metrics (e.g., proof size or structural properties) are either unavailable before construction or do not reliably reflect modeling difficulty.
Therefore, we adopt the requirement count as a deterministic and practical partition criterion.

\noindent \textbf{Metrics.}
In this work, we design three metrics to evaluate the consistency and correctness of the constructed formal models.

\paragraph{Consistency}
A formal model is consistent if it is not contradictory to itself.
We evaluate the consistency level of a model through the Proof Obligation Discharge Rate (PDR).
For a model $M$ with a set of proof obligations $PO_M$, the PDR is defined as follows:
{\footnotesize \[
PDR = \frac{\sum_{po \in PO_M} \mathbb{I}[Discharged(M,po)]}{|PO_M|}
\]}
The $PDR$ metric corresponds to the second conjunct in \autoref{eq:correctness}, measuring the proportion of proof obligations that are successfully discharged.
In our experiment, we report the $PDR$ of the final model after refinement and repair, since the final model's consistency reflects the consistency of the entire refinement chain.

\paragraph{Correctness}
A model $M$ is correct if it covers and fulfills all requirements from the requirement document $REQ_M$.
A requirement $req \in REQ_M$ is considered covered if it is presented in $M$ without errors, and $req$ is fulfilled if:
\begin{enumerate}
    \item It is covered in $M$, i.e. $Covered(M,req)$
    \item POs associated with $req$ are discharged, i.e. $\bigwedge_{po \in PO_{req}} Discharged(M, po)$
\end{enumerate}
During model construction, we instruct the corresponding LLM to generate labels for elements corresponding to the requirements they represent.
In this way, we can compute both requirement coverage (RC) and requirement fulfillment (RF):

{\footnotesize \[
\begin{aligned}
RC = \frac{\sum_{req \in REQ_M} \mathbb{I}[Covered(M,req)]}{|REQ_M|}, \quad
RF = \frac{\sum_{req \in REQ_M} \mathbb{I}[Covered(M,req) \land (\bigwedge_{po \in PO_{req}} Discharged(M, po))]}{|REQ_M|}
\end{aligned}
\]
}
$RC$ and $RF$ are computed across multiple refinement layers, since the refine layer hides the preserved requirements and only includes new requirements or updated requirements.

\paragraph{Validity of Assumption}
When computing the metrics, we assume that once a requirement is covered and fulfilled in an abstract model $M_i$, it remains so in the subsequent refinement models $M_{i+1}, \dots, M_n$.
This assumption holds provided the refinement is correct, i.e., all refinement POs are discharged.
In practice, a failed refinement PO does not necessarily imply that all requirements from $M_i$ are violated, but it is also unclear which subset of requirements may be affected.
To make $RC$ and $RF$ computable across refinement layers, we therefore adopt this assumption, and validate it empirically by reporting the $PDR$ for refinement-related POs for all refinement layers.

\subsection{RQ1. Overall Performance} \label{sec:rq1}
We assess the effectiveness of Event-B Agent for end-to-end formal model synthesis and repair by comparing it with the three baseline methods introduced in Section~\ref{sec:evalsetting}.
\autoref{tab:RQ1} summarizes the performance of all four methods across datasets partitioned by system complexity, as defined in Section~\ref{sec:evalsetting}, and reports their overall results.
The methods under comparison are:
\begin{itemize}
    \item \textbf{LLM + auto provers}: a naive baseline that synthesizes models using an LLM and attempts to discharge the proofs with SMT solvers and theorem provers.
    \item \textbf{Cursor}: a general-purpose coding agent baseline implemented with Cursor.
    \item \textbf{PAT-Agent}: a baseline for formal model synthesis without exploiting proof information.
    \item \textbf{Event-B Agent}: our proposed framework.
\end{itemize}
The results demonstrate that our method consistently outperforms the three baselines on the metrics $PDR$, $RC$, and $RF$, both overall and across different levels of system complexity.

\noindent
\textbf{Consistency.}
Event-B Agent attains an overall $PDR$ of 97.86\%, indicating that the synthesized models are largely self-consistent, with only 2.14\% of proof obligations remaining undischarged. 
In contrast, the LLM + auto provers, Cursor, and PAT-Agent baselines achieves $PDR$s of 89.20\%, 90.07\%, and 95.56\% respectively, falling short of our method and exhibiting greater variance. 
Notably, Cursor’s $PDR$ fluctuates by 12.5\% across different complexity levels, which makes it the most unstable method among all.
Without a structured framework, its performance solely relies on the capability of the underlying LLM.
While PAT-Agent attains stronger performance, its $PDR$ still varies more than ours and declines as system complexity increases. 
In contrast, Event-B Agent consistently maintains $PDR$ above 97.0\% across all dataset partitions.

\noindent
\textbf{Correctness w.r.t. Requirements.}
Metrics $RC$ and $RF$ measure how well a formal model covers and satisfies system requirements. 
As discussed in Section~\ref{sec:evalsetting}, this requires validating the assumption of refinement correctness. 
The $PDR$ for refinement-specific proof obligations (Refinement $PDR$) achieved by Event-B Agent is reported in \autoref{tab:RQ2_refine}. 
The full version of Event-B Agent reaches a Refinement $PDR$ of 92.56\%, indicating that the assumption holds for most refined models. 
On this basis, $RC$ and $RF$ can be used as approximations of requirement-level correctness.  

The last two columns of \autoref{tab:RQ1} show that Event-B Agent achieves overall $RC$ and $RF$ values of 97.13\% and 93.79\%, respectively, which are the highest among all evaluated methods. 
In particular, $RC$ exceeds the second-best method by 4.63\% and $RF$ by 18.01\%.
Across all dataset partitions, our method consistently delivers the best performance with lower variance on both metrics, demonstrating that models produced by Event-B Agent cover and satisfy the largest proportion of system requirements. 
It is also worth noting that lower $RC$ and $RF$ values imply that $PDR$ could have been superficially inflated, as the generated formal system may bypass more challenging requirements.
Moreover, the ratio $RF/RC$ for our method is 0.97, substantially higher than those of the three baselines (0.75, 0.77, and 0.82). 
This indicates that once a requirement is captured in the formal model, Event-B Agent is able to discharge nearly all corresponding proof obligations. 
We therefore conclude that Event-B Agent performs the best in terms of correctness with respect to system requirements.

\begin{table*}[h!]
\footnotesize
\centering
\begin{tabular}{llccc}

\toprule
System Complexity & Method & PDR & RC & RF \\
\midrule
\multirow{4}{*}{Simple}
 & LLM + auto provers & 0.8615 & 0.8889 & 0.7546 \\
 & Cursor & 0.8418 & 0.9278 & 0.6370 \\
 & PAT Agent & 0.9783 & 0.9167 & 0.7806 \\
 & Event-B Agent & \textbf{0.9954} & \textbf{0.9639} & \textbf{0.9417} \\

\midrule
\multirow{4}{*}{Medium}
 & LLM + auto provers & 0.9492 & 0.8974 & 0.7090 \\
 & Cursor & 0.9668 & 0.9057 & 0.8114 \\
 & PAT Agent & 0.9653 & 0.8623 & 0.7572 \\
 & Event-B Agent & \textbf{0.9700} & \textbf{0.9667} & \textbf{0.9347} \\

\midrule
\multirow{4}{*}{Complex}
 & LLM + auto provers & 0.8653 & \textbf{0.9886} & 0.6051 \\
 & Cursor & 0.8936 & 0.8450 & 0.6175 \\
 & PAT Agent & 0.9232 & 0.9825 & 0.7357 \\
 & Event-B Agent & \textbf{0.9706} & 0.9834 & \textbf{0.9373} \\

\midrule
\multirow{4}{*}{Overall}
 & LLM + auto provers & 0.8920 & 0.9250 & 0.6896 \\
 & Cursor & 0.9007 & 0.8928 & 0.6886 \\
 & PAT Agent & 0.9556 & 0.9205 & 0.7578 \\
 & Event-B Agent & \textbf{0.9786} & \textbf{0.9713} & \textbf{0.9379} \\
\bottomrule

\end{tabular}
\caption{Overall performance of Event-B Agent compared to baselines with different dataset complexities.
GPT-5 with medium reasoning level is used as the backbond LLM for all methods.}
\label{tab:RQ1}
\end{table*}

\subsection{RQ2. Ablation Study}

The two key components of Event-B Agent are the refinement strategy planning step and the model \& proof repair step.
The latter integrates formal verification with a novel repair guidance system, namely, repair rules recommendation, fix strategy decision LLM, and atomic repair function execution.
As shown in Section~\ref{sec:rq1}, Event-B Agent outperforms the baselines, which do not exploit proof information for repairing.
To address RQ2, we perform a finer-grained evaluation regarding the model \& proof repair step.
Specifically, we examine the performance of Event-B Agent when proof information is available, but repair guidance is disabled, leaving the LLM to directly repair the model based on the proof state.

\noindent
\textbf{Ablation Baselines.}
The ablation study evaluates Event-B Agent against three variants:
\begin{enumerate}
    \item \textbf{None Enabled}: Both refinement and repair guidance are removed. This baseline isolates the capability of the LLM in model construction and repair when provided only with the model and verification results from the model checker and theorem provers.  
    \item \textbf{Refinement only}: Refinement is retained, but repair guidance is removed. Since refinement introduces additional proof obligations and helps to reduce proving efforts at the same time, this baseline assesses how refinement affects the overall performance.
    \item \textbf{Repair guidance only}: Refinement is omitted while repair guidance is retained. This baseline isolates the effectiveness of the repair guidance component.  
\end{enumerate}

\noindent
\textbf{Results and Discussion.}
\autoref{tab:RQ2} summarizes the ablation results. 
Across all metrics, the full version of Event-B Agent consistently delivers the strongest overall performance, outperforming the ablation baselines. 
On the ``Simple'' and ``Medium'' partitions, some ablations obtain marginally higher scores, but these gains are limited and do not persist across the dataset as a whole.
For $RC$ and $RF$, the overall performance of the full version of Event-B Agent again surpasses all baselines by a significant margin, confirming the pattern observed in Section~\ref{sec:rq1}. 
Since these metrics reflect requirement-level correctness, lower coverage may unexpectedly inflate the observed $PDR$ values.
Taken together, these results highlight the complementary contributions of refinement and repair guidance to the effectiveness of Event-B Agent.

\begin{table*}[h!]
\footnotesize
\centering

\begin{minipage}[c]{0.58\columnwidth}

\begin{tabular}{llccc}

\toprule
Dataset & Method & PDR & RC & RF \\
\midrule
\multirow{4}{*}{Simple}
 & (1) None enabled & 0.9279 & 0.7611 & 0.6833 \\
 & (2) Refinement only & 0.9425 & 0.8944 & 0.8296 \\
 & (3) Repair guidance only & 0.9890 & \textbf{0.9861} & \textbf{0.9583} \\
 & (4) Event-B Agent & \textbf{0.9954} & 0.9639 & 0.9417 \\

\midrule
\multirow{4}{*}{Medium}
 & (1) None enabled & 0.9799 & 0.8316 & 0.8316 \\
 & (2) Refinement only & \textbf{0.9861} & 0.8970 & 0.8543 \\
 & (3) Repair guidance only & 0.9659 & 0.9632 & 0.8863 \\
 & (4) Event-B Agent & 0.9700 & \textbf{0.9667} & \textbf{0.9347} \\

\midrule
\multirow{4}{*}{Complex}
 & (1) None enabled & 0.9600 & 0.9162 & 0.7954 \\
 & (2) Refinement only & 0.9664 & 0.8951 & 0.8210 \\
 & (3) Repair guidance only & 0.9529 & 0.8990 & 0.7548 \\
 & (4) Event-B Agent & \textbf{0.9706} & \textbf{0.9834} & \textbf{0.9373} \\

\midrule
\multirow{4}{*}{Overall}
 & (1) None enabled & 0.9559 & 0.8363 & 0.7701 \\
 & (2) Refinement only & 0.9650 & 0.8955 & 0.8350 \\
 & (3) Repair guidance only & 0.9693 & 0.9494 & 0.8665 \\
 & (4) Event-B Agent & \textbf{0.9786} & \textbf{0.9713} & \textbf{0.9379} \\
\bottomrule

\end{tabular}
\caption{Ablation studies on the refinement strategy planning and model \& proof repair components of Event-B Agent.}
\label{tab:RQ2}

\end{minipage}
\hfill
\begin{minipage}[c]{0.37\columnwidth}

\begin{tabular}{ll p{0.22\columnwidth}}
\toprule
Dataset & Method & Refinement PDR \\
\midrule
\multirow{2}{*}{Simple} & Refinement only & 0.7778 \\
 & Event-B Agent & \textbf{1.000} \\
\midrule
\multirow{2}{*}{Medium} & Refinement only & 0.7878 \\
 & Event-B Agent & \textbf{0.9556} \\
\midrule
\multirow{2}{*}{Complex} & Refinement only & 0.4653 \\
 & Event-B Agent & \textbf{0.8213} \\
\midrule
\multirow{2}{*}{Overall} & Refinement only & 0.6769 \\
 & Event-B Agent & \textbf{0.9256} \\
\bottomrule
\end{tabular}

\caption{Refinement PO Discharge Rate (Refinement $PDR$) of Event-B Agent and the refinement-related ablation baseline.
Refinement $PDR$ estimates the confidence that the assumption of refinement correctness holds.}

\label{tab:RQ2_refine}
    
\end{minipage}

\end{table*}

While comparing baseline 2 (``Refinement only'') and Event-B Agent with their counterparts without refinement (baseline 1, ``None enabled'', and baseline 3, ``Repair guidance only''), we generally observe a positive trend across all metrics when refinement is present. 
There are, however, two notable exceptions. 
First, in the ``Simple'' partition, although Event-B Agent achieves a higher $PDR$ than baseline 3, its $RC$ and $RF$ values are slightly lower. 
As system complexity increases, the $RC$ and $RF$ values for baseline 3 decline, and Event-B Agent surpasses it. 
This suggests that refinement becomes increasingly important for ensuring correctness in more complex systems.
Second, in the ``Complex'' partition, baseline 2 achieves only slightly better $RF$ than baseline 1, and a lower $RC$. 
From \autoref{tab:RQ2_refine}, we observe that the ``Refinement only'' baseline attains a Refinement $PDR$ of just 46.53\% in this partition, compared to 77.78\% and 78.78\% in the other two. 
This correlation between low Refinement $PDR$ and reduced performance is consistent with the explanation in Section~\ref{sec:evalsetting}: without sufficiently validated refinement steps, the observed requirement-level metrics become inflated and unreliable.
Overall, these results confirm that high Refinement $PDR$ is essential for validating refinement links and preserving the inductive property in Event-B.

Regarding the repair guidance component, we observe that the ``Repair guidance only'' baseline performs significantly better than baseline 1 in the ``Simple'' partition, demonstrating its effectiveness in less complex systems. 
However, its $PDR$ decreases as system complexity increases, likely due to implementation limitations.
The repair module currently supports only a fixed set of atomic repair functions, which suffices for simple cases but is inadequate for more complex obligations.
When combined with refinement, this limitation is mitigated.
Refinement decomposes complex systems into smaller ones, in which proofs fall within the reach of the repair guidance system, underscoring the complementary nature of the two components.
This synergy is reflected in the Event-B Agent achieving the highest $PDR$ in two partitions and in the overall results.

In summary, RQ2 shows that refinement and repair guidance are mutually reinforcing. 
Refinement reduces the complexity of individual proofs by decomposing model construction into smaller steps, making them more amenable to repair. 
Conversely, effective model and proof repairs improve the discharge of refinement obligations, enabling the construction of higher-quality models that provably preserve the requirements established in simpler abstract models.

\subsection{RQ3. Efficiency}

\begin{table*}[h!]
\footnotesize
\centering
\begin{tabular}{p{0.06\linewidth} p{0.035\linewidth} p{0.035\linewidth} l p{0.035\linewidth} p{0.035\linewidth} l p{0.035\linewidth} p{0.035\linewidth} l p{0.035\linewidth} p{0.035\linewidth} l}

\toprule
Dataset &
\multicolumn{3}{l}{\begin{tabular}[c]{@{}l@{}}
Refinement Strategy \\
Planning
\end{tabular}} &
\multicolumn{3}{l}{Model Synthesis} &
\multicolumn{3}{l}{Model \& Proof Repair} &
\multicolumn{3}{l}{Overall} \\

\cmidrule(lr){2-4} \cmidrule(lr){5-7} \cmidrule(lr){8-10} \cmidrule(lr){11-13}
 & Time & \#Calls & \#Tokens & Time & \#Calls & \#Tokens & Time & \#Calls & \#Tokens & Time & \#Calls & \#Tokens\\

\midrule
Simple & 
1.22 & 1.00 & 4973.00 &
19.75 & 13.55 & 79359.56 &
15.76 & 15.55 & 127694.89 &
37.00 & 30.11 & 212027.44 \\

\midrule
Medium & 
1.20 & 1.00 & 4940.33 &
25.48 & 13.33 & 97841.11 &
52.89 & 53.22 & 2393013.44 &
83.21 & 67.56 & 2495794.89 \\

\midrule
Complex &
1.18 & 1.00 & 6131.22 &
29.97 & 13.89 & 121982.78 &
62.46 & 59.45 & 2137659.11 &
103.14 & 74.33 & 2265773.11 \\

\midrule
Overall &
1.20 & 1.00 & 5348.19 & 
25.07 & 13.59 & 99727.81 &
43.71 & 42.74 & 1552789.15 &
74.45 & 57.33 & 1657865.15 \\
\bottomrule

\end{tabular}

\caption{The efficiency of each Event-B Agent component and the overall framework.
Time is the execution time in minutes, \#Calls is the number of LLM calls, and \#Tokens is the number of tokens consumed.}
\label{tab:efficiency}
\end{table*}

\noindent
We evaluate the efficiency of Event-B Agent using three metrics: Time, the total execution time (in minutes) per component and overall, \#Calls, the number of LLM invocations, and \#Tokens, the number of tokens consumed.
\autoref{tab:efficiency} reports their distributions across components and the overall model construction cost.
Time and \#Calls for refinement strategy planning remain stable across system complexities, as it runs once per system.
Variation in \#Tokens likely reflects differences in requirement counts.
For model synthesis, \#Calls are similar across partitions, while Time and \#Tokens increase with system complexity, indicating higher per-call cost for complex systems.
For model \& proof repair, both Time and \#Calls increase with complexity.
Notably, \#Tokens for ``Complex'' is lower than ``Medium'', consistent with our observation that refinement reduces proving effort.
Although the “Complex” partition has more POs, they are not necessarily harder to discharge.

The model \& proof repair step takes on average \SI{43.71}{minutes}, \SI{42.74}{LLM} calls, and 1552789.15 tokens, largely due to high number of proof obligations (182.41 on average).
\#Calls is lower than PO count because many POs are discharged automatically, while the remaining require more complex reasoning or additional premises, where LLM-based model \& proof repair becomes necessary.
In practice, Event-B experts typically spend substantially more time on interactive proofs.
In contrast, the average time to attempt discharging a single PO in Event-B Agent remains below \SI{0.30}{minutes} across all partitions (0.18, 0.30, and \SI{0.22}{minutes}), with an overall average of \SI{0.24}{minutes}.
Overall, this suggests that the repair overhead scales primarily with the number of POs, rather than exhibiting uncontrolled growth with system complexity.

\subsection{RQ4. How does Event-B Agent construct and repair models?}



\begin{figure}[t]
\centering

\begin{minipage}[c]{0.33\columnwidth}
\includegraphics[width=1\columnwidth]{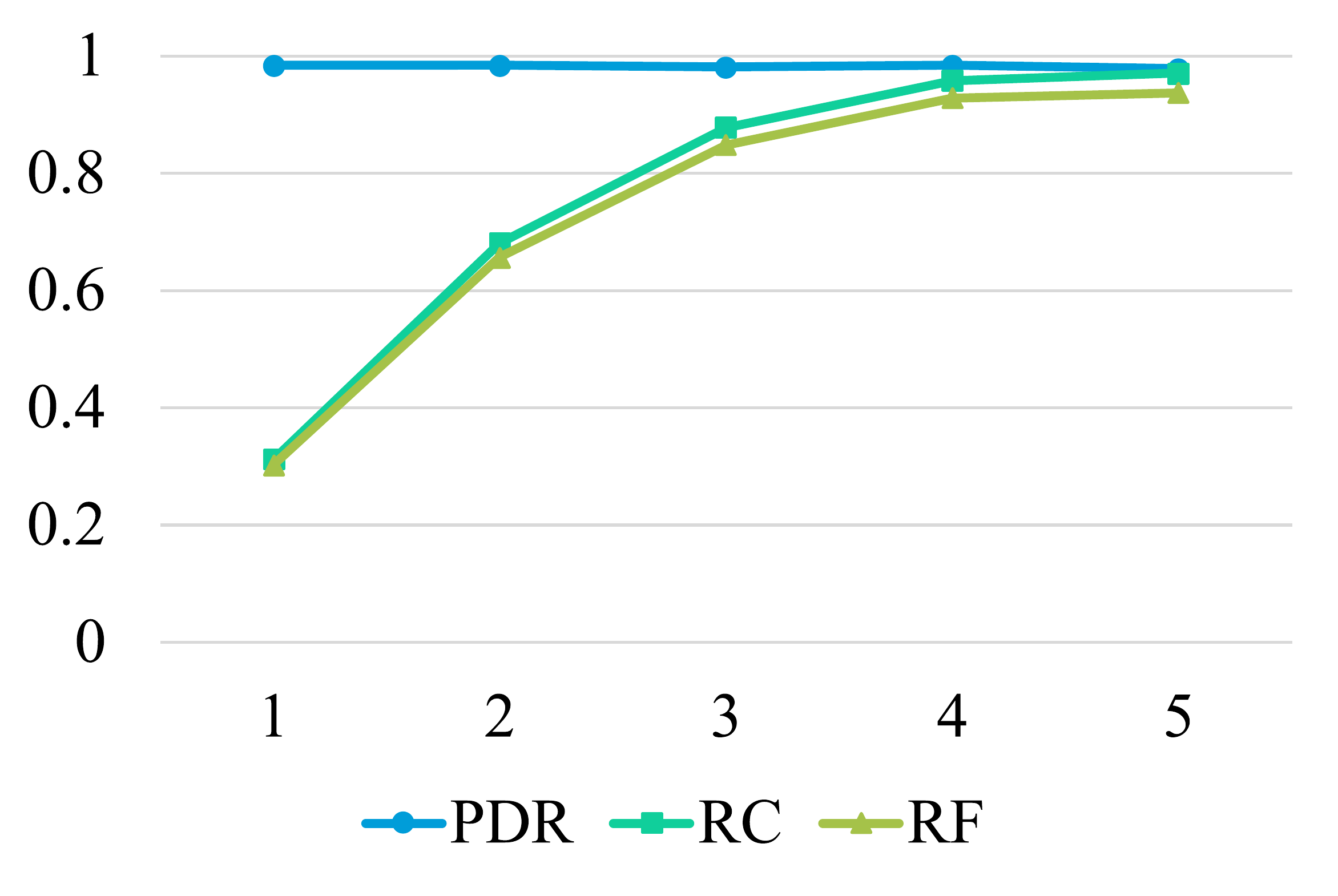}
\caption{Evolution of the three metrics across refinement steps.}
\label{fig:refinement}
\end{minipage}
\hfill
\begin{minipage}[c]{0.65\columnwidth}
\centering
\includegraphics[width=1\columnwidth]{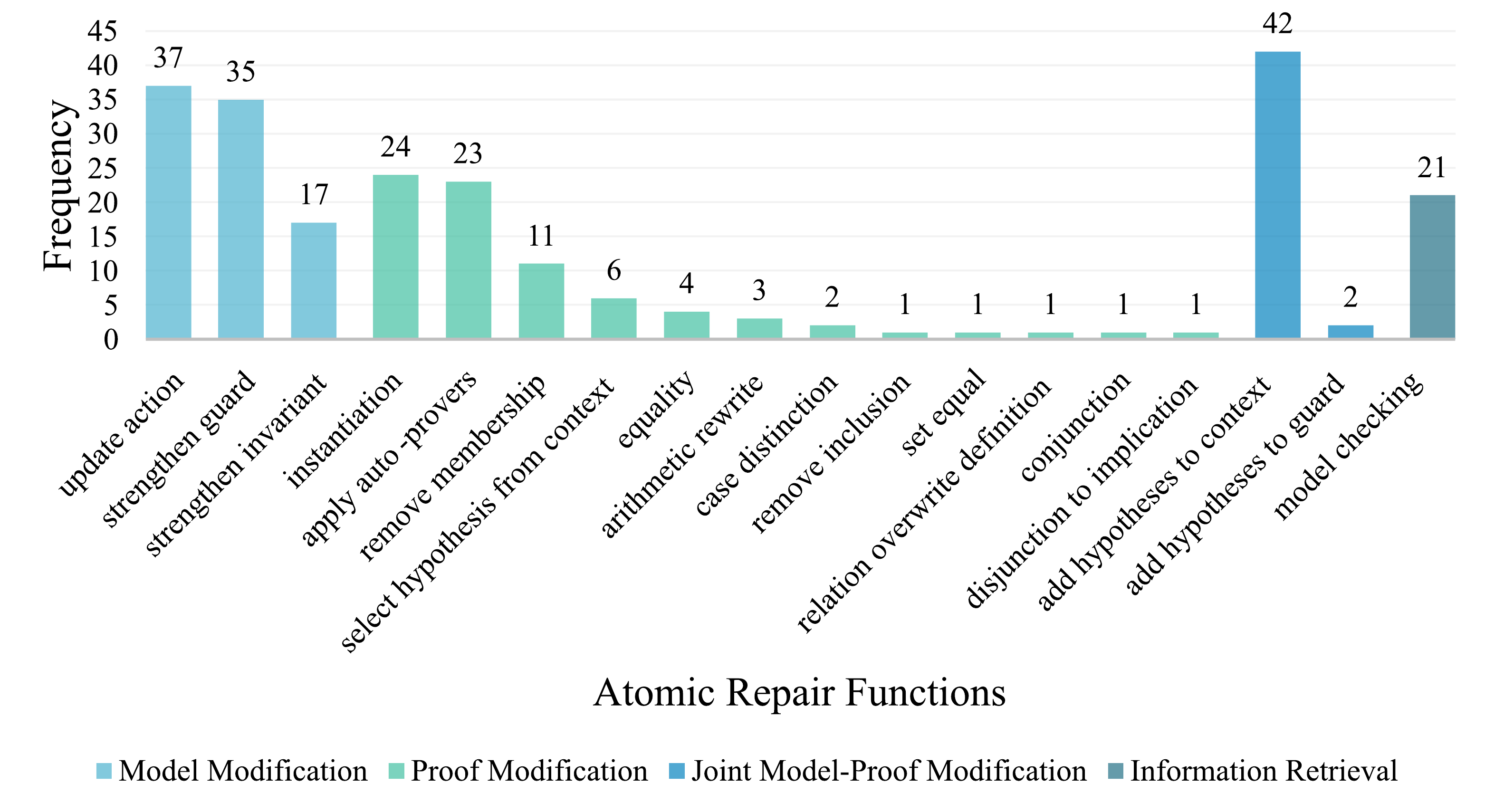}
\caption{Distribution of model and proof repair functions contributing to successful PO discharge.}
\label{fig:distribution}
\end{minipage}

\end{figure}

\noindent
\textbf{Refinement.}
In \autoref{fig:refinement}, we present how the three metrics evolve as refinement proceeds.
Across all refinement steps, the average $PDR$ is maintained at a consistently high level (97.86–98.50\%). 
The average $RC$ gradually increases from 31.19\% in the abstract models to 97.13\% in the final model, while average $RF$ rises from 30.24\% to 93.79\%.
This trend shows that refinement effectively expands requirement coverage and fulfillment without compromising proof obligation discharge rate.

\noindent
\textbf{Atomic Repair Functions.}
In \autoref{fig:distribution}, we show the frequency of atomic repair functions invoked by Event-B Agent that successfully contributed to discharging proof obligations.  
Overall, functions from all four categories are involved, demonstrating their effectiveness. 
Among them, model modifications account for 38.36\% of successful invocations, followed by proof modifications at 33.62\%. 
Joint model–proof modifications contribute 18.97\%, showing the impact of operations that simultaneously affect both the model and the proof. 
Finally, retrieve information functions, such as model checking, make up 9.10\%, indicating that even non-modifying actions play a role in guiding future repairs. 
This distribution shows that while proof- and model-level operations dominate individually, the hybrid category provides a substantial contribution.

\noindent
\textbf{Qualitative Evaluation of Repair.}
In this section, we present three representative repairs from the motivating example (\autoref{fig:running_example}) to illustrate how Event-B Agent uses POs to guide model repair.

\textbf{(1) Ensuring Well-Definedness.}
Initially, the invariant $FUN$--1 generated by Event-B Agent is $has\_j=TRUE \Rightarrow f(j)=min(ran(f))$, which is not well defined because $j$ could be out of the domain of $f$.
The agent then enters the repair step and calls $StrengthenInvariant$ on this invariant, and updates it to be $has\_j=TRUE \land j \in dom(f) \Rightarrow f(j)=min(ran(f))$.
After this repair to the model, the well-definedness PO for this invariant is successfully discharged.

\textbf{(2) Ensuring Refinement Correctness.}
In the abstract machine $M_{abstract}$, the event $stop\_empty$ doesn't assign a value to $j$, while the corresponding event in the concrete machine $M_{concrete}$ assigns $j:=0$.
As a result, a PO has been generated to ensure that under guard $n=0$ of this event, $j$ must be 0.
This PO cannot be discharged due to missing hypotheses.
Event-B Agent repairs this by calling the $StrengthenInvariant$ function, and adds invariant $n=0 \Rightarrow j=0$ to $M_{concrete}$, which helps to discharge the generated PO.
Furthermore, invariant preservation POs are generated for this new invariant, and are all discharged in $M_{concrete}$, guaranteeing that the repair is consistent and correct.

\textbf{(3) Invariant Preservation.}
As discussed in Section~\ref{sec:repair}, Event-B Agent is capable of ensuring that every event preserves the invariants.
In the motivating example, Event-B Agent calls the $StrengthenGuard$ function and adds $grd7: 1 \leq i$ to event $find\_not\_better$, since $i$ can never be 0 due to the design of the algorithm.
As a result, the post-state invariant $inv'$ becomes vacuously true and the invariant preservation PO is discharged.

\section{Related Work}

\subsection{Autoformalization and Repair}

LLMs have been explored to reduce the effort of translating informal natural language requirements into formal specifications~\cite{zhang2025position}.
One direction is property synthesis, where natural language is mapped to LTL~\cite{cosler2023nl2spec, fuggitti2023nl2ltl}, though this covers only a fragment of specification languages and cannot capture full system models.
Other work addresses program specification generation~\cite{wen2024enchanting, wu2023lemur} or proof synthesis and repair~\cite{first2023baldur, lu2024proof, carrott2024coqpyt}, but these remain task-specific rather than providing an integrated modeling and verification workflow.

More recent studies target system-level models, such as Alloy~\cite{hong2025effectiveness} and Event-B~\cite{capozucca2025ai}, but these approaches are limited in scope and rely on direct text-to-code mappings.
The closest work is PAT-Agent~\cite{zuo2025pat}, which synthesizes PAT models and validates them via model checking.
While effective for complex systems and requirements, its verification is confined to full-state-space model checking.
By contrast, Event-B Agent combines bounded model checking with theorem proving, enabling counterexample-guided repair and proof-based correctness.

Beyond LLMs, the B and Event-B community has advanced construction and refinement methods~\cite{alkhammash2015building,dupont2021event,mashkoor2017refinement}, model repair~\cite{cai2019automatic,cai2022fast,kobayashi2024repairing}, and code generation~\cite{mery2011automatic,furst2014code,rivera2017code}.
These efforts remain human-guided, repair-specific, or downstream, whereas Event-B Agent unifies synthesis, repair, and verification in an automated framework.


\subsection{Neurosymbolic Methods}
Neurosymbolic methods \cite{chaudhuri2021neurosymbolic} combine the expressiveness of neural learning with the verifiability of symbolic reasoning.
Recent directions include minimal neural activation patterns for verifiable robustness~\cite{geng2024learning}, automata embeddings for reinforcement learning with semantic guarantees~\cite{yalcinkaya2025provably}, and specification learning from combined membership and preference queries~\cite{shah2023learning}.
Other efforts integrate logic into differentiable frameworks, such as tensorized LTL\textsubscript{f} constraints verified in Isabelle/HOL and utilized for neural training~\cite{chevallier2025formally}, or world models that follow principles for physical interpretability, such as structuring latent spaces by physical intent and partitioning outputs for verifiability~\cite{peper2025four}.

Event-B Agent shares this neurosymbolic philosophy but targets formal verification, using LLMs for semantic tasks while ensuring correctness through model checking and theorem proving in a unified synthesis-and-repair loop.

\subsection{Proof Tactic Recommendation}

Tactic recommendation in interactive theorem proving has long aimed to reduce user effort. Early approaches learned from existing proofs and replayed strategies, including PaMpeR for Isabelle/HOL~\cite{nagashima2018pamper}, TacticToe for HOL4~\cite{gauthier2021tactictoe}, Tactician for Coq~\cite{blaauwbroek2020tactician}, and PSL, a strategy language that expands into concrete tactic sequences~\cite{nagashima2017proof}.

Recent work leverages transformers and LLMs. MagnusHammer~\cite{mikula2023magnushammer} enhances premise selection with contrastive learning, while LeanDojo~\cite{yang2023leandojo} combines LLMs with large proof libraries in a retrieval-augmented prover. Other directions explore proof-step prediction~\cite{gloeckle2023temperature}, recursive proof generation~\cite{wang2024proving}, lifelong agents such as LeanAgent~\cite{kumarappan2024leanagent}, and approaches such as Lean-STaR~\cite{lin2024lean}, which interleave chain-of-thought reasoning with tactic prediction to improve proof success.

These efforts center on tactic prediction and retrieval within provers, whereas Event-B Agent embeds tactic guidance in a broader neurosymbolic framework, where model construction, verification, and repair progress together rather than as isolated tasks.


\section{Threats to Validity}

\subsection{Internal Validity}

\noindent\textbf{Potential Biases and Assumptions of Requirements.}
Although our evaluation annotates requirements with lightweight labels (e.g., ``EQP''), Event-B Agent can operate on unstructured natural language.
The labels are used solely to compute RC and RF, require minimal annotation effort, and do not encode modeling structure.
Since all methods use the same requirements and specification language, any bias from requirement phrasing affects them uniformly.

We also assume internally consistent requirements, whereas real-world specifications may be inconsistent.
Handling such cases (e.g., supporting human-in-the-loop) is left for future work.


\noindent\textbf{Requirement Capturing.}
To compute RC and RF, formal model elements are labeled with requirement identifiers during synthesis.
In our experiments, this labeling is performed by LLMs but occasionally violates the expected format, affecting automatic matching and underestimating RC and RF.
We manually corrected such cases for all methods to ensure fair and consistent evaluation.

\noindent \textbf{Prompt Sensitivity and Reproducibility.}
LLM-based systems are sensitive to prompt phrasing.
However, in Event-B Agent, the prompt structure is fixed for each task, with only task-specific information programmatically injected.
More importantly, system behavior is governed by PO-guarded verification, and every LLM-proposed change is checked by deterministic tools (the Event-B compiler, model checker, and theorem provers), limiting the impact of minor prompt variations.

Although GPT-5 introduces nondeterminism through stochastic decoding, we mitigate this via schema-based constraints and verifier-mediated acceptance.
Moreover, reproducibility here refers to the stability of the improvement process rather than identical modeling outputs, as multiple correct models may exist.
The PO-guarded repair ensures that accepted changes do not regress the model with respect to the target PO, and our ablation results empirically demonstrate the robustness of the pipeline to LLM nondeterminism.

\subsection{External Validity}
\noindent \textbf{Scalability.}
Since theorem proving is unbounded in principle, there is no theoretical limit on the system size supported.
Although LLMs impose context-window constraints, these were not encountered in our experiments.
For larger systems, refinement and decomposition can limit prompt scope by operating on partial models while preserving correctness via POs.


\noindent\textbf{Implementation Limitations.}
The current libraries of repair rules and fix strategies are not exhaustive, as they were derived empirically from proof patterns across diverse systems.
Extending these libraries to cover additional rules, proof states, and repair functions is left for future work.

\section{Conclusion}

We presented Event-B Agent, a novel framework for end-to-end formal model construction that integrates refinement-based development with coordinated model construction and proof derivation.
Evaluation on 27 formal systems demonstrates consistent improvements over baselines in consistency and correctness, with ablation studies highlighting the complementary contributions of refinement and repair while maintaining reasonable efficiency.
Future work includes expanding the library of repair rules and fix strategies, and supporting human-in-the-loop mechanisms to address requirement inconsistencies.
Overall, this work takes a step toward practical and sound LLM-assisted formal development guided by the principle of correctness by construction.

\section{Data Availability}
The implementation of Event-B Agent and the datasets used are publicly available\footnote{\url{https://github.com/HongshuW/EventB_Agent}, \url{https://doi.org/10.5281/zenodo.19642103}}.

\bibliographystyle{ACM-Reference-Format}
\bibliography{reference}


\end{document}